# Electrical detection of the inverse Edelstein effect on the surface of $SmB_6$

Jehyun Kim[1], Chaun Jang[2], Xiangfeng Wang[3], Johnpierre Paglione[3], Seokmin Hong[2],

Shehrin Sayed[4], Dongwon Chun[5] and Dohun Kim[1*]

[1]Department of Physics and Astronomy, and Institute of Applied Physics, Seoul National University, Seoul 08826, Korea

[2]Center for Spintronics, Korea Institute of Science and Technology, Seoul 02792, Korea

[3]Maryland Quantum Materials Center, Department of Physics, University of Maryland, College Park, MD 20742, USA

[4]Electrical Engineering and Computer Science, University of California, Berkeley, California 94720, USA

[5]Advanced Analysis Center, Korea Institute of Science and Technology, Seoul 02792, Korea

*Corresponding author: dohunkim@snu.ac.kr

ABSTRACT: We report the measurement of spin current-induced charge accumulation, the inverse Edelstein effect (IEE), on the surface of single-crystal candidate topological Kondo insulator $SmB_6$. The robust surface conduction channel of $SmB_6$ has been shown to exhibit a large degree of spin-momentum locking, and the spin-polarized current through an external ferromagnetic contact induces spin-dependent charge accumulation on the surface of $SmB_6$. The dependence of the IEE signal on the bias current, an external magnetic field direction, and temperature are consistent with an anticlockwise spin texture of the $SmB_6$ surface band in momentum space. The direction and magnitude of this effect, compared with the normal Edelstein signal, are clearly explained by the Onsager reciprocal relation. Furthermore, we estimate the spin-to-charge conversion efficiency, i.e., the IEE length, to be 4.46 nm, which is an order of magnitude larger than the efficiency found in other typical Rashba interfaces, implying that the Rashba contribution to the IEE signal may be small. Building upon existing reports on the surface charge and spin conduction nature of this material, our results provide additional evidence that the surface of $SmB_6$ supports a spin-polarized conduction channel.

## I. INTRODUCTION

Three-dimensional (3D) topological insulators (TIs) are a newly developed class of insulators with a bulk band gap in which time-reversal symmetry-protected metallic surface states reside. The spin-momentum locking exhibited by the surface conduction channels make TIs a promising platform for exploring new physics such as Majorana quasi-particle states, and for applications in various spintronic devices [1–3]. However, in conventional 3D TIs, the Fermi level naturally resides in the bulk conduction or valence bands owing to unintentional doping, resulting in the hindrance of surface-driven phenomena by bulk carriers [4–6]. Recently, $SmB_6$, a Kondo insulator, has been predicted to be a member of a newly classified family of strong TIs, topological Kondo insulators (TKIs), in which the topologically protected surface states reside in the bulk Kondo band gap at low temperatures and the Fermi level is guaranteed to be inside the bulk gap [7–9]. A large degree of current-induced spin polarization on the surface of $SmB_6$ as well as robust surface conduction have been demonstrated in various experiments [10–19], implying that $SmB_6$ is a strong candidate for TKIs free from bulk effects.

Here, we report an additional demonstration that the surface of $SmB_6$ indeed exhibits transport phenomena consistent with a spin-momentum locked spin texture through observation of the spin current-induced charge accumulation, the inverse Edelstein effect (IEE). Distinct from a previous spin injection experiment using microwave-induced spin pumping on the surface of $SmB_6$ [11], we use a near-DC electrical method to generate charge accumulation through the IEE. Charge accumulation can be generated not only by injecting spin-polarized current generated using a ferromagnetic metal into the surface of $SmB_6$, but also by extraction of the spin-polarized current generated from the surface of $SmB_6$. The resultant charge accumulation is measured as the voltage difference between two nonmagnetic contacts on the surface. The measured dependence of the charge accumulation on the direction and magnitude of the bias current, the external magnetic field direction, and the temperature are all consistent with the spin-momentum locking of the $SmB_6$ surface state.

## II. MATERIALS AND METHODS

### A. Material growth

Single crystals of $SmB_6$ were grown with Al flux in the ratio of $SmB_6$ : Al = 1 : 200–250 starting from elemental Sm and B with a stoichiometry of 1 to 6. The initial materials were placed in an alumina crucible and loaded in a tube furnace under an Ar atmosphere. The assembly was heated to 1250–1400 °C and maintained at that temperature for 70–120 h, then cooled at −2 °C/ h to 600–900 °C, followed by faster cooling. The SmB$_6$ samples were placed in sodium hydroxide to remove the residual Al flux.

### B. Device fabrication

A 2 nm thick Al layer was deposited on the polished (100) surface of SmB$_6$ by electron beam evaporation followed by oxidation on a hotplate under ambient conditions. The resulting thin Al oxide layer prevents direct contact of the ferromagnetic metal with SmB$_6$ and acts as a tunnel barrier between SmB$_6$ and the ferromagnetic metal. This generally enhances the spin injection and detection ratio by alleviating the conductance mismatch problem [20] (see Supplementary Information S1 [21]). Standard e-beam lithography was used to fabricate the electrodes. A permalloy (Py) layer with a lateral size of 150 × 150 μm$^2$ and thickness of 20 nm was used as a ferromagnetic spin source for spin injection and extraction. The layer was capped with 15 nm of Au using electron beam evaporation. Non-ferromagnetic contacts used for the source, drain, and voltage probes were formed by e-beam lithography patterning and Al oxide etching with a buffered oxide etchant followed by the deposition of 5 nm Ti/ 80 nm Au using electron beam evaporation. To avoid the direct wire bonding to the Py layer which can damage the properties of the Py layer, the Au electrode acting as the wire bonding pad for the contact with the ferromagnetic metal was made using electron beam evaporation and an additional insulating layer was made below this Au electrode by overdosing electron beam on electron beam resist (PMMA 950A6) with a dose of 10000 μC/cm$^2$, which enables the Au electrode to be connected directly to the Py layer, not to SmB$_6$ surface [see the inset to Fig. 1(a)].

### C. Transport measurements

The device was placed in a commercial variable temperature cryostat (Quantum Design PPMS) for low-temperature electrical measurements. For all the electrical measurements, standard lock-in-based four-point probe measurements were performed. An AC current was applied through the interfacial tunnel oxide between the ferromagnetic metal Py and the $SmB_6$ surface using an AC current source (Keithley 6221), and a lock-in amplifier (Stanford Research Systems SR830) was used to detect the voltage difference between two Au contacts.

## III. RESULTS

### A. Principle of electrical measurement for IEE

The Edelstein effect is one of the well-known effects involving charge-to-spin conversion intimately related to the spin Hall effect. In materials with spin-momentum locking, the flow of a charge current produces non-equilibrium spin polarization through the Edelstein effect [22]. The Onsager reciprocal effect of the Edelstein effect is called the IEE, where a non-equilibrium spin accumulation in a two-dimensional electron gas generates charge accumulation perpendicular to its spin direction [23,24].

To detect the charge accumulation on the surface of $SmB_6$ arising from the IEE, a Py layer is used as a spin source to induce non-equilibrium spin accumulation on the $SmB_6$ surface. Figure 1(a) shows the electrical measurement configuration for the IEE, where a bias current $I_b$ flows through Py on the $SmB_6$ parallel to the y axis, and the transverse voltage difference $V_{yx}$, defined as $V_+ - V_-$, is measured between two nonmagnetic Au contacts positioned at the ends of $SmB_6$ on the x axis while sweeping an external magnetic field along the y axis. The measured $V_{yx}$ can be classified into four cases according to the directions of $I_b$ and the Py magnetization (***M***).

Figures 1(b) – (e) show schematic top views of the device illustrated in Fig. 1(a), and the charge accumulation along the x axis due to the IEE where the accumulated spin-up (spin-down) electrons are depicted with red (blue) arrows parallel (anti-parallel) to the +y direction and the direction of the accumulation is shown by the grey arrows pointing to the Au voltage probes. We note that the same amount of electrons with spin anti-parallel to the accumulated spin, in a steady state, must flow in the direction toward the center of the device owing to the zero net current along the x axis in an open circuit condition. Moreover, because the direction of the Py majority spin is opposite to that of its magnetization and the majority spin of Py is mainly coupled to the $SmB_6$ surface channel, the contact resistance between Py and the spin-up channel of $SmB_6$ is larger (smaller) than that between Py and the spin-down channel of $SmB_6$ for $M$ parallel to the +y (-y) direction [25,26].

In the case of injection ($I_b$ parallel to the -y direction) in which spin-polarized electrons are injected into the surface of $SmB_6$, for $M$ parallel to the +y (-y) direction, more spin-down (spin-up) electrons are accumulated on the surface of $SmB_6$. The accumulated electrons subsequently have net momentum in the -x (+x) direction due to the spin-momentum locking, resulting in a higher electrochemical potential at the left (right) side and, eventually, $V_{yx} <$ (>) 0. On the other hand, for extraction ($I_b$ parallel to the +y direction), where the spin-polarized electrons from $SmB_6$ are extracted and tunnel into Py, for $M$ parallel to the +y (-y) direction, more spin-up (spin-down) electrons are left behind on the surface of $SmB_6$ owing to the high contact resistance. These electrons subsequently have net momentum in the +x (-x) direction due to the spin-momentum locking, resulting in a higher electrochemical potential at the right (left) side and, eventually, $V_{yx} >$ (<) 0. We confirmed the non-zero $V_{yx}$ induced by spin current injection/extraction using a simulation based on the semi-classical model for charge and spin transport (see Supplementary

Information S2 [21] and Ref. [25]). In summary, the expected behaviors of $V_{yx}$ as functions of the external magnetic field $H_y$ for injection and extraction are described in Fig. 1(f) and 1(g), respectively, where $H_c$ is the switching field of Py and the IEE signal $\Delta V_{yx}$, defined as $V_{yx}$ (***M*** // +y) – $V_{yx}$ (***M*** // -y), is negative (positive) for injection (extraction). We also define the polarities of the hysteresis loops in Fig. 1(f) and 1(g) as negative and positive, respectively.

### B. Electrical measurement of the IEE signal and Onsager reciprocal relation

We first report the expected behavior of the aforementioned IEE signal on the surface of $SmB_6$, which is reflected in a non-zero $\Delta V_{yx}$. As shown in Fig. 2(a), $V_{yx}$ is measured by sweeping an external magnetic field along the y axis to control the magnetization direction of Py while applying $I_b$ along the y axis. Figures 2(b) and 2(c) show representative $V_{yx}$ values as functions of $H_y$ recorded with $I_b$ of +150 μA and −150 μA, respectively, at 1.8 K. For $I_b$ of +150 μA (−150 μA), the spin extraction (injection) results in a hysteresis loop with a positive (negative) polarity in agreement with our expectation, which is clearly consistent with the anticlockwise spin-momentum relation in $SmB_6$ (see Supplementary Information S3 [21] and Ref. [27,28]). The $\Delta V_{yx}$ extracted from the hysteresis loops under different $I_b$ values exhibits a linear response to $I_b$, as shown in Fig. 2(d), implying that the current-induced spin injection and extraction lead to a non-zero $\Delta V_{yx}$.

The IEE signal can also be analyzed quantitatively using the Onsager reciprocal relation. The Onsager reciprocal relation is a universal relation for any setup in the linear response regime. It states that the ratio of the measured voltage to the bias current does not change even when the voltage and current terminals are exchanged [26,29]. However, when a time-reversal symmetry breaking field such as ***M*** is present, the sign of the field should be reversed in the reciprocity relation. Thus, the Onsager reciprocal relation is given by

$$\frac{V_{12}(\boldsymbol{M})}{I_{34}} = \frac{V_{34}(-\boldsymbol{M})}{I_{12}}, \qquad (1)$$

where $V_{ab}$ is defined as $V_a - V_b$ and $I_{cd}$ denotes the current that flows from terminal c to terminal d. Figure 3 clearly exhibits the Onsager reciprocal relation expressed by Eq. (1) between the potentiometric spin measurement, where the ferromagnetic metal is used as the spin detector, and its reciprocal measurement for the IEE, where $SmB_6$ is used as the spin detector. Figures 3(a) and 3(c) respectively show a schematic drawing of the potentiometric spin measurement and the corresponding $V_{34}$ recorded with $I_{12}$ of 100 μA at 1.8 K while sweeping an external magnetic field along the y axis. Figures 3(b) and 3(d) respectively show a schematic drawing of the IEE measurement and the corresponding $V_{12}$ recorded with $I_{34}$ of 100 μA at 1.8 K while sweeping an external magnetic field along the y axis. The results show a hysteresis loop with negative polarity for the potentiometric measurement and a loop with positive polarity for the IEE measurement consistent with the Onsager reciprocal relation Eq. (1). More specifically, in the potentiometric spin measurement, the spin voltage $\Delta V_{34}$ can be expressed as [30]

$$\frac{\Delta V_{34}}{I_{12}} = \frac{V_{34}(\boldsymbol{M}) - V_{34}(-\boldsymbol{M})}{I_{12}} = R_B P_{FM}(\boldsymbol{p} \cdot \boldsymbol{M_u}), \qquad (2)$$

where $\Delta V_{34}$ is the reciprocal value of $\Delta V_{12}$ in this study. $\Delta V_{34}$ is proportional to the bias current $I_{12}$, ballistic channel resistance $R_B$, ferromagnetic metal spin polarization $P_{FM}$, and the inner product between the surface channel spin polarization $\boldsymbol{p}$ under a positive bias current and the unit vector along the ferromagnetic metal magnetization $\boldsymbol{M_u}$. Eq. (1) and Eq. (2) can be combined to yield

$$\frac{\Delta V_{12}}{I_{34}} = \frac{V_{12}(\boldsymbol{M}) - V_{12}(-\boldsymbol{M})}{I_{34}} = -R_B P_{FM}(\boldsymbol{p} \cdot \boldsymbol{M_u}), \qquad (3)$$

where the negative sign is due to the Onsager reciprocal relation. As expected in Eq. (3), the slope from the linear fitting shown in Fig. 2(d) has an opposite sign to that of the bias current dependence of the spin voltage [19,31]. Furthermore, the magnitude of the slope in Fig. 2(d) is 2.7 mΩ, which is slightly larger than the 2.3 mΩ previously reported for a potentiometric geometry experiment [19]. This difference can be attributed to the non-linearity of the contact resistance between the ferromagnetic metal and $SmB_6$. As the IEE signal follows the Onsager reciprocal relation, we estimate $|p|$ of $SmB_6$ to be 27% from both the inverse and normal Edelstein effect results (see Supplementary Information S4 [21] and Ref. [10,19,32]). Therefore, the IEE signal $\Delta V_{12}$ electrically measured through both spin injection and extraction supports the conclusion that $SmB_6$ indeed has an anticlockwise surface spin texture in momentum space.

### C. Magnetization orientation dependence of IEE signal

To further confirm the spin-momentum relation, we study how the IEE signal depends on the magnetization orientation. Figures 4(a) and 4(b) show the schematic top views of the measurement configurations when an external magnetic field is applied along the y axis and x axis, respectively, under an $I_b$ of 100 μA at 1.8 K. The corresponding results are shown in Fig. 4(c) and 4(d). Owing to the anticlockwise spin texture of the $SmB_6$ surface band, charge accumulation by spin-to-charge conversion on the surface of $SmB_6$ occurs along the x axis as depicted in Fig. 4(a), resulting in a measurable $\Delta V_{yx}$, as shown in Fig. 4(c). On the other hand, we can predict that charge accumulation occurs along the y axis when $M$ is parallel to the x axis, as depicted in Fig. 4(b), resulting in no voltage difference between the two voltage probes at high positive or negative $H_x$, as shown in Fig. 4(d). The intermittent non-zero signal in Fig. 4(d) is likely to be due to magnetic domains in which the transient magnetizations have some y-axis components in the process of magnetization reversal through domain wall motion [33]. Furthermore, the result in Fig. 4(d) also excludes the possibility that the measured $\Delta V_{yx}$ originates from spurious effects such as the Hall effect, where non-zero $\Delta V_{yx}$ can arise independently of the magnetization orientation owing to the fringe field of the ferromagnetic injector [34]. We also confirm that SmB$_6$ with all Au contacts in which the Py layer is replaced by the Au layer shows no such field dependent $V_{yx}$, demonstrating that the Py layer has crucial role in spin injection/extraction (see Supplementary Information S5). Therefore, the magnetization dependence of the IEE signal further offers the conclusion that the measured $\Delta V_{yx}$ clearly reflects the anticlockwise spin texture of the SmB$_6$ surface band.

### D. Temperature dependence of IEE signal

The surface origin of $\Delta V_{yx}$ was examined by investigating the temperature dependence of $\Delta V_{yx}$. As shown in Fig. 5(a), the temperature-dependent electrical resistance $R(T)$ of SmB$_6$ diverges from 12 to 4 K, exhibiting thermally activated behavior, and starts to saturate at 4 K, exhibiting surface-dominated transport properties as confirmed previously [16–18]. Fig. 5(b) shows $V_{yx}$ as functions of $H_y$ under $I_b$ of +100 μA at temperatures ranging from 4.5 to 1.8 K (marked by red dots in Fig. 5(a)). The variation of the IEE signal $\Delta V_{yx}$ with the measurement temperature is extracted from Fig. 5(b) and summarized in Fig. 5(c). As the temperature increases, $\Delta V_{yx}$ constantly decreases and vanishes at around 4 K. This resembles the temperature dependence behavior of the SmB$_6$ electrical resistance, which shows a crossover from surface to bulk-dominated charge conduction at around 4 K. Moreover, although SmB$_6$ is a heavy metal where the spin Hall and inverse spin Hall effects can occur, the signal from the inverse spin Hall effect does not contribute to the measured $\Delta V_{yx}$ at elevated temperatures. This may be largely attributed to the thick bulk channel of SmB$_6$ in which the reduced spatially averaged spin current in a thicker spin detector material diminishes the inverse spin Hall signal [35]. We also note that the temperature dependence of the measured $\Delta V_{yx}$ exhibits a similar behavior to the results of previous temperature-dependent $\Delta V_{xy}$ in the potentiometric measurement configuration. This confirms that the Onsager reciprocal relation is valid at different temperatures [19]. Therefore, the temperature dependence of the measured $\Delta V_{yx}$ gives strong support for the measured $\Delta V_{yx}$ as having originated from the surface states of SmB$_6$, and largely excludes bulk effects such as the inverse spin Hall effect.

## IV. DISCUSSION

The pinning of the Fermi energy near the hybridization-induced gap due to the hybridization of localized *f* electrons with conduction electrons ensures surface-dominated transport in SmB$_6$ at low temperatures [17,18], thereby excluding the possibility that bulk effects such as the inverse spin Hall effect might have contributed to the measured IEE signal. However, the IEE signal can arise from both the Rashba surface states and topologically-protected surface states because spin-momentum locking is present in both types of surface states. Although it is difficult to separately measure the contributions of the Rashba and topological surfaces to the IEE signal, the measured IEE signal is very likely to consist mainly of the signal from the topological surface. We arrive at this conclusion through the analysis of the IEE length, $\lambda_{IEE}$, which is the spin-to-charge conversion efficiency given by [25,36]

$$j_C = \lambda_{IEE} j_S, \; \lambda_{IEE} = \frac{|p|\lambda}{\pi}, \tag{4}$$

where the charge current density in A m$^{-1}$, $j_c$, and spin current density in A m$^{-2}$, $j_s$, are connected through $\lambda_{IEE}$, which is proportional to the absolute value of the spin polarization $|p|$ and the mean free path of the channel $\lambda$. In SmB$_6$, because the surface conduction is mainly contributed by β band electrons, $\lambda$ for the β band (52 nm) [37] and the spin polarization (27%) are used to obtain $\lambda_{IEE}$. In our case, $\lambda_{IEE}$ is 4.46 nm, which is comparable to the $\lambda_{IEE}$ found in α-Sn film topological insulators without bulk effects [38]. It is an order of magnitude larger than the $\lambda_{IEE}$ found in various other Rashba interfaces with typical values of 0.1–0.4 nm owing to the compensation between the two Fermi contours of Rashba interfaces [39–43]. This implies that the Rashba contribution to the IEE signal should be small. Moreover, a large $\lambda_{IEE}$ also indicates that $SmB_6$ is a promising candidate for spintronic devices that are potentially useful as efficient spin sources and detectors. With the recently developed technique of increasing the temperature range of surface-dominated transport in $SmB_6$ by applying strain [44], the material also has potential in spintronic applications at elevated temperatures. Our observation presents a route for the potential application of $SmB_6$ both in fundamental investigations of the interplay between nontrivial topology and electron correlation, and in applied spin transport physics in strongly correlated systems.

## ACKNOWLEDGMENT

This research was supported by the Basic Science Research Program (Grant No. NRF-2015R1C1A1A02037430 and 2018R1A2A3075438) through the National Research Foundation of Korea (NRF) funded by the Ministry of Science, ICT and Future Planning, and the Creative-Pioneering Researchers Program through Seoul National University (SNU). Research at the University of Maryland was supported by the Gordon and Betty Moore Foundation's EPiQS Initiative through Grant No. GBMF9071 and the Maryland Quantum Materials Center. The electrical measurements used shared facilities funded by the KIST Institutional Program and the National Research Council of Science & Technology (NST) grant (No. CAP-16-01-KIST). J. K. and C. J. contributed equally to this work.

**Figure captions**

FIG. 1. Principle of electrical measurement of the Inverse Edelstein Effect (IEE). (a) Schematic of the measurement setup and anticlockwise spin texture of the surface band in $SmB_6$ near the Fermi energy. Inset: Optical microscope image of the device. The length of the white scale bar is 100 μm. (b) – (e) Schematic top views of the charge accumulation due to the IEE in the cases of ***M*** // *+y* under injection (b), ***M*** // *+y* under extraction (c), ***M*** // *-y* under injection (d), and ***M*** // *-y* under extraction (e). The grey arrows represent the direction in which electrons with spin-up or spin-down move. (f), (g) The expected inverse Edelstein signals for spin injection (f) and extraction (g).

FIG. 2. Electrical measurement of the IEE signal. (a) Schematic of the electrical measurement configuration. A bias current $I_b$ is applied along the y axis, and the voltage difference is measured between two Au contacts while sweeping a magnetic field along the y axis. (b), (c) The measured

$V_{yx}$ as a function of the y component of an external magnetic field $H_y$ for $I_b$ of +150 μA (b) and −150 μA (c). (d) Dependence of the IEE signal $\Delta V_{yx}$ as a function of $I_b$ measured at 1.8 K.

FIG. 3. The Onsager reciprocal relation. (a), (b) Schematic measurement setup for the potentiometric spin measurement (a) and its reciprocal measurement for the IEE (b). (c) $V_{34}$, defined as $V_3 − V_4$, as a function of an external magnetic field swept along the y axis under $I_{12}$ of 100 μA at 1.8 K, measured in Fig. 3(a) configuration [19]. (d) $V_{12}$, defined as $V_1 − V_2$, as a function of an external magnetic field swept along the y axis under $I_{34}$ of 100 μA at 1.8 K, measured in Fig. 3(b) configuration.

FIG. 4. Magnetization orientation dependence of the IEE signal. (a), (b) Schematic top view of the measurement configuration. An external magnetic field is swept along the y axis in (a) and along the x axis in (b). The grey arrows represent the direction in which electrons with spin-up or spin-down move. (c), (d) $V_{yx}$ as a function of an external magnetic field swept along the y axis (c) and along the x axis (d) under $I_b$ of +100 μA at 1.8 K. The magnetization $M$ is parallel to the y axis (parallel to current direction) in (c) and parallel to the x axis (perpendicular to current direction) in (d).

FIG. 5. Temperature dependence of the IEE signal. (a) Electrical resistance of $SmB_6$ as a function of temperature under a bias current of 300 μA. (b) $V_{yx}$ measured by sweeping an external magnetic field parallel to the y axis under a bias current of +100 μA at different temperatures ranging from 1.8 to 4.5 K. Each curve is offset by 1 μV for clarity. (c) The IEE signal $\Delta V_{yx}$ extracted from Fig. 4(b) as a function of temperature.

Figure 1

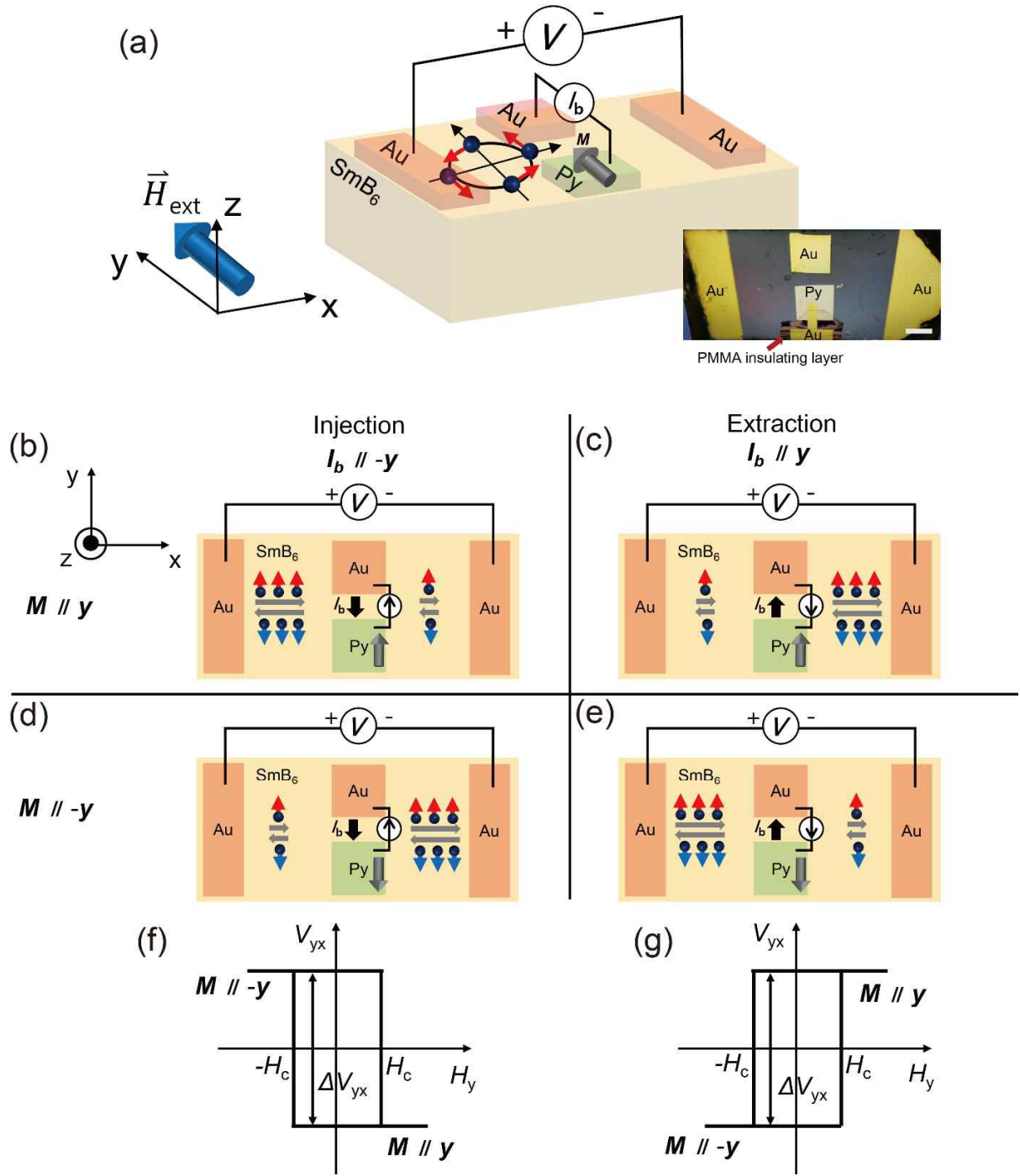

Figure 2

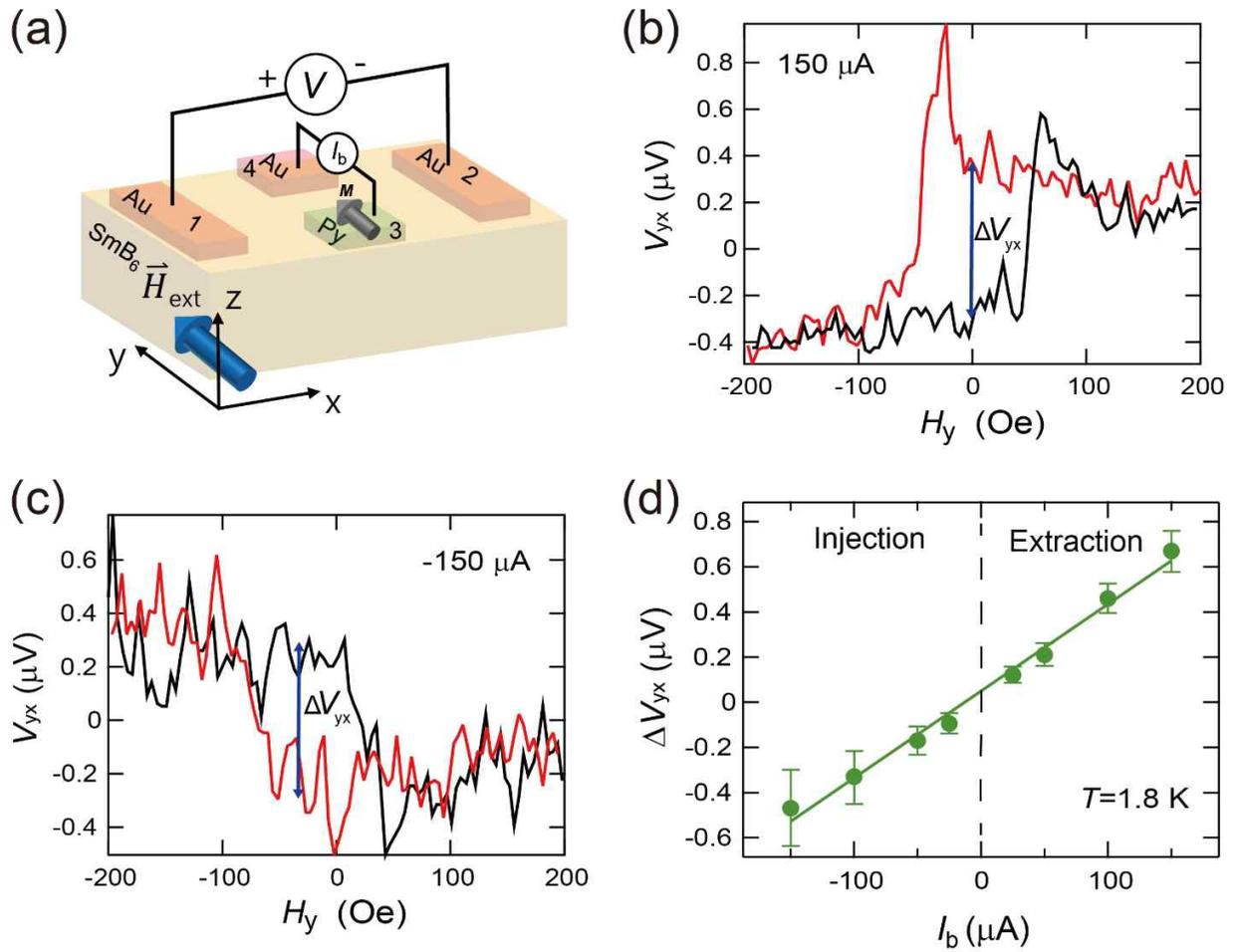

Figure 3

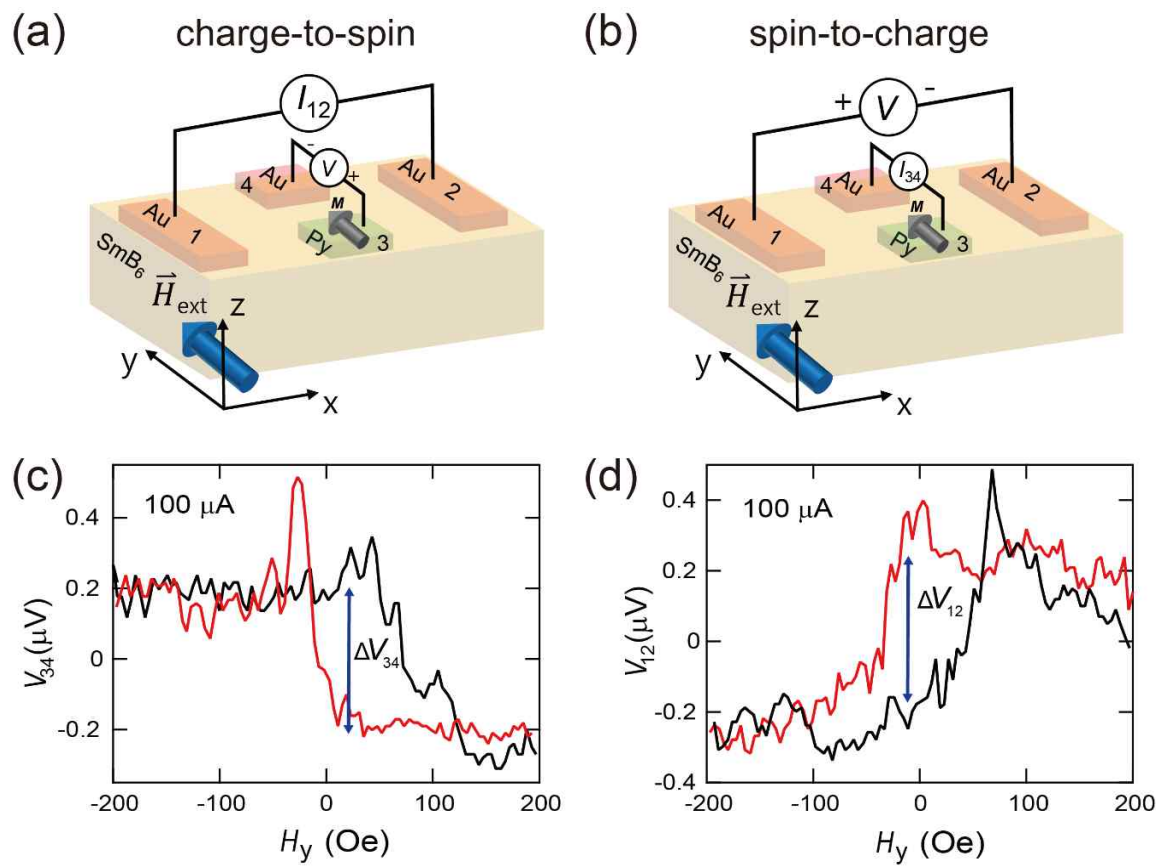

Figure 4

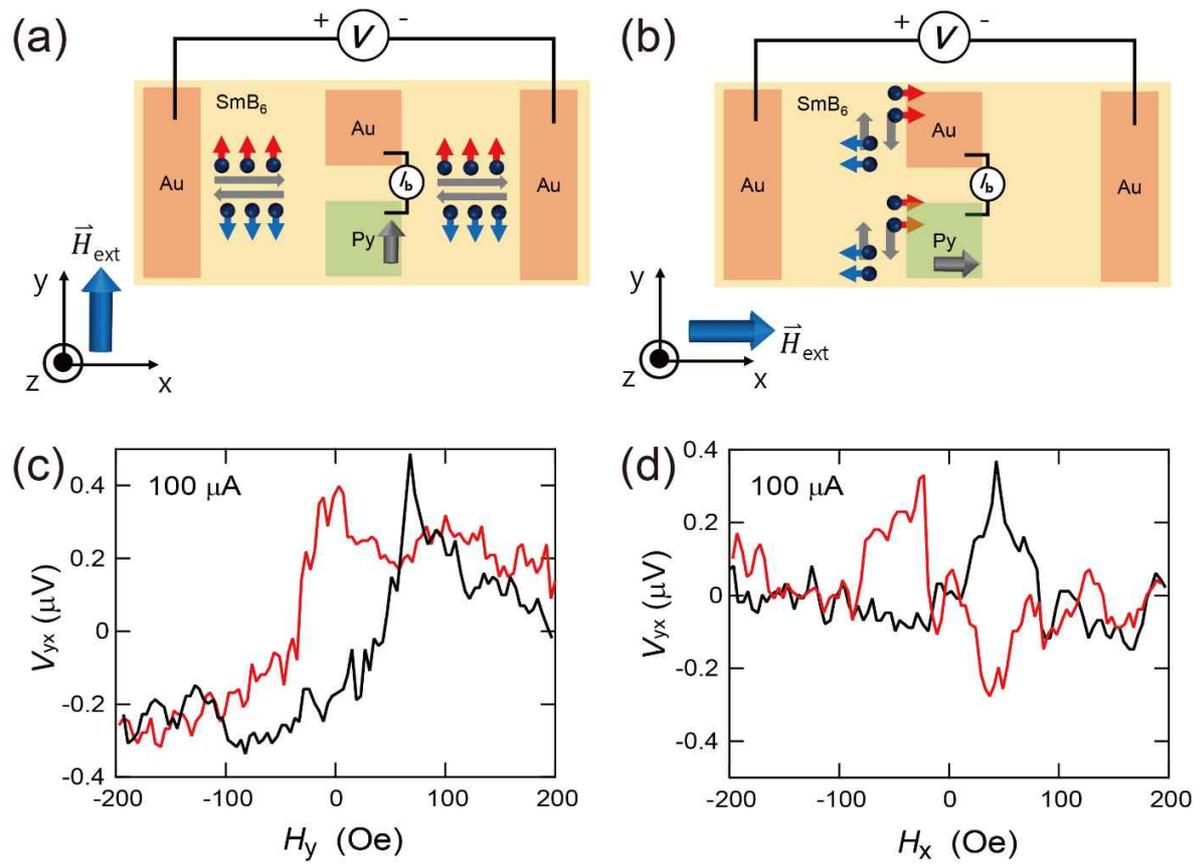

Figure 5

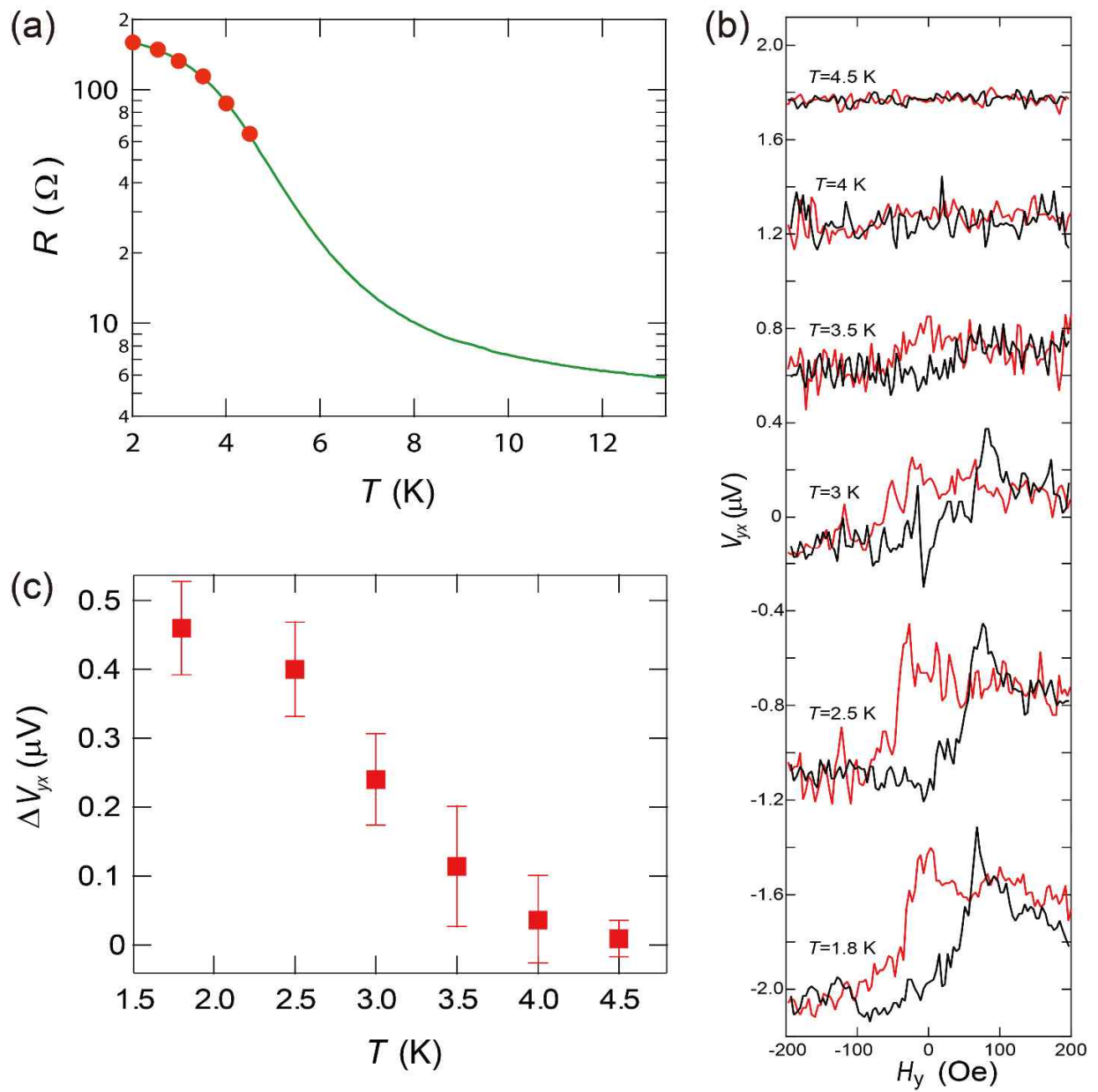